\documentclass[preprint,preprintnumbers, prd, floatfix, superscriptaddress,nofootinbib] {revtex4-1}

\usepackage{epsfig}
\usepackage{subfigure}
\usepackage{dcolumn}
\usepackage{bm}
\usepackage[usenames ,dvipsnames]{xcolor}
\usepackage{slashed}
\usepackage{graphicx,color}
\usepackage{hyperref}

\begin{document}
\title{$CP$ violation in singly Cabibbo suppressed $D\to \pi a_0(980)$ decays}

\author{Yu-Kuo Hsiao}
\email{yukuohsiao@gmail.com}
\affiliation{School of Physics and Electronic Engineering, Shanxi Normal University, 
Taiyuan 030031, China}

\author{Shu-Ting Cai}
\email{18734581917@163.com}
\affiliation{School of Physics and Electronic Engineering,
Shanxi Normal University, Taiyuan 030031, China}

\author{Yan-Li~Wang}
\email{ylwang0726@163.com}
\affiliation{School of Physics and Electronic Engineering, Shanxi Normal University,
Taiyuan 030031, China}

\date{\today}

\begin{abstract}
The singly Cabibbo suppressed (SCS) decays $D\to \pi a_0$, with $a_0\equiv a_0(980)$, 
have been measured with the branching-fraction ratios 
$r^{+/-}_{\rm ex}\equiv {\cal B}(D^0\to\pi^- a_0^+)/{\cal B}(D^0\to\pi^+ a_0^-)=7.5^{+2.5}_{-0.8}\pm 1.7$ and 
$r^{+/0}_{\rm ex}\equiv {\cal B}(D^+\to\pi^0 a_0^+)/{\cal B}(D^+\to\pi^+ a_0^0)=2.6\pm 0.6\pm 0.3$,
deviating significantly from the short-distance expectations $(r^{+/-},r^{+/0})\simeq (0.07,0.2)$.
This discrepancy indicates the necessity of long-distance rescattering effects.
In particular, the process $D\to K^*K\to a_0\pi$ generates ${\cal M}_s$ comparable in magnitude to 
${\cal M}_d$ in the amplitude ${\cal M}=\lambda_d{\cal M}_d+\lambda_s{\cal M}_s$, with 
$\lambda_q\equiv V_{cq}^*V_{uq}$, accompanied by nontrivial strong phases 
essential for $CP$ violation. Consequently, 
the direct $CP$ asymmetries naturally arise at the ${\cal O}(10^{-3})$ level, for example, 
${\cal A}_{CP}(D^0\to\pi^- a_0^+)=(-0.7\pm 0.1\pm 0.1\pm 0.1)\times 10^{-3}$,
${\cal A}_{CP}(D^+\to\pi^0 a_0^+)=(-1.4\pm 0.1\pm 0.1\pm 0.1)\times 10^{-3}$, and
${\cal A}_{CP}(D^0\to\pi^+ a_0^-)=(-2.1\pm 0.9\pm 1.1\pm 0.4)\times 10^{-3}$.
These results establish SCS $D\to \pi a_0$ decays as a new avenue for probing $CP$ violation.
\end{abstract}

\maketitle
\section{introduction}
To account for the matter--antimatter extreme asymmetry of the universe,
$CP$ violation has been proposed as one of the essential Sakharov conditions~\cite{Sakharov:1967dj}. 
While it has been well established in strange- and beauty-hadron weak decays, 
its presence in the charm sector remained elusive until the first observation 
of a nonzero difference in the time-integrated $CP$ asymmetries, defined as
$\Delta {\cal A}_{CP}\equiv{\cal A}_{CP}(D^0\to \pi^+\pi^-)
-{\cal A}_{CP}(D^0\to K^+ K^-)
=(-1.54\pm 0.29)\times 10^{-3}$~\cite{LHCb:2019hro}. 
Whether the measured $\Delta {\cal A}_{CP}$ can be fully accommodated 
within the Standard Model (SM)~\cite{Cheng:2012wr,Bhattacharya:2012ah,Li:2012cfa,
Franco:2012ck,Brod:2012ud,Khodjamirian:2017zdu,Chala:2019fdb,Grossman:2019xcj,
Buccella:2019kpn,Cheng:2019ggx,Schacht:2021jaz,Bediaga:2022sxw,Pich:2023kim,Cheng:2024hdo} 
or instead points to possible new physics contributions~\cite{Grossman:2006jg,
Hochberg:2011ru,Chen:2012am,Feldmann:2012js,Dery:2019ysp,Iguro:2024uuw,
Sinha:2025cuo,Fleischer:2025zhl} remains an open question~\cite{Muller:2015rna}. 
Addressing this issue has induced extensive searches for ${\cal A}_{CP}$.
However, no signal with a significance comparable to $\Delta {\cal A}_{CP}$ 
has yet been observed in the $D$ decays~\cite{LHCb:2022lry,LHCb:2025ezf,
Belle:2025cub,Belle-II:2025rmf,Belle-II:2025wsy, 
LHCb:2024rkp,LHCb:2024jpt,Belle-II:2025zqj,LHCb:2025zgk}. 

Currently, the BESIII Collaboration has reported the first observation 
of the singly Cabibbo-suppressed (SCS) process $D\to \pi a_0(980)$, 
followed by the resonant decay $a_0(980)\to\pi\eta$~\cite{BESIII:2024tpv}.  
Defining $a_0\equiv a_0(980)$, the measured ratios of the branching fractions are 
\begin{eqnarray}\label{data1}
&&
(r^{+/-}_{\rm ex},r^{+/0}_{\rm ex})\equiv\bigg(
\frac{{\cal B}(D^0\to\pi^- a_0^+)}{{\cal B}(D^0\to\pi^+  a_0^-)},
\frac{{\cal B}(D^+\to\pi^0  a_0^+)}{{\cal B}(D^+\to\pi^+ a_0^0)}\bigg)
=(7.5^{+2.5}_{-0.8}\pm 1.7,2.6\pm 0.6\pm 0.3)\,,
\end{eqnarray}
where the common resonant contribution from $a_0\to\pi\eta$ has been factored out. 
Within the factorization approach~\cite{Cheng:2022vbw}, the short-distance (SD) 
$W$-emission contributions to $D\to \pi a_0$ are estimated as 
$r^{+/-}_{\rm SD}=0.065$ and $r^{+/0}_{\rm SD}=0.16$, 
which are smaller than the experimental values by about two 
and one orders of magnitude, respectively. 
Similar discrepancies are also observed in Cabibbo-allowed $D_s^+$ decays,
where the measured branching fractions of 
$D_s^+\to\pi^+\omega$~\cite{LHCb:2022pjv}, 
$D_s^+\to\pi^{0}a_0^{+}(\pi^{+}a_0^{0})$~\cite{BESIII:2019jjr}, 
and $D_s^+\to\rho^0 a_0^{+}$~\cite{BESIII:2021aza}
exceed theoretical estimates based solely on SD contributions by 
one to two orders of magnitude. Including LD rescattering effects 
significantly alleviates these discrepancies~\cite{Fajfer:2003ag,
Hsiao:2019ait,Ling:2021qzl,Yu:2021euw}. 

Motivated by these successful resolutions,
we incorporate LD rescattering effects in the SCS $D\to\pi a_0$ decays. 
In particular, rescattering processes induced by
the quark-level weak transitions $c\to d u\bar d$ and $c\to s u\bar s$
contribute to ${\cal M}_d$ and ${\cal M}_s$, respectively, 
in the amplitude ${\cal M}=\lambda_d {\cal M}_d+\lambda_s {\cal M}_s$, 
with $\lambda_q=V_{cq}^*V_{uq}$. Since both ${\cal M}_d$ and ${\cal M}_s$, 
as well as their relative strong phase, are expected to be sizable, 
the resulting direct $CP$ asymmetry can naturally reach a level comparable to the observed 
$\Delta {\cal A}_{CP}$. 
We therefore propose a systematic analysis of the SCS $D\to\pi a_0$ decays, 
aiming to account for the measured branching fractions
and to provide the first predictions for $CP$ asymmetries in these modes.

\section{Formalism}
The external and internal $W$-emission ($W_{\rm em}$) processes
as the SD contributions to the SDS $D\to \pi a_0$ decays 
correspond to the $T$ and $C$ amplitudes, respectively. 
Within the factorization framework,
the amplitudes $T$ and $C$ can be expressed as~\cite{Cheng:2022vbw}
\begin{eqnarray}\label{TandC}
&&
{T}_{\rm S}(D^0\to\pi^- a_0^{+})
=ia_1(m_D^2-m_{\pi}^2)f_{a_0^+}F^{D^0\to \pi^-}\,,\nonumber\\ 
&&
{T}_{\rm P}(D^0\to\pi^+ a_0^{-})
=ia_1(m_D^2-m_{a_0}^2)f_{\pi^+} F^{D^0\to a_0^-}\,,\nonumber\\ 
&&
{C}_{\rm P}(D^+\to\pi^0 a_0^+)
=ia_2(m_D^2-m_{a_0}^2)f_{\pi^0} F^{D^+\to a_0^+}\,,\nonumber\\ 
&&
{T}_{\rm P}(D^+\to\pi^+ a_0^{0})
=ia_1(m_D^2-m_{a_0}^2)f_{\pi^+}F^{D^+\to a_0^0}\,.
\end{eqnarray}
Here, the effective parameters $a_{1,2}$ arise from the factorization framework,
the decay constants $(f_{a_0},f_\pi)$ correspond to the vacuum-produced mesons, and
the form factors $F^{D\to (a_0,\pi)}$ parameterize the $D$-meson transition matrix elements.
The subscripts $S$ and $P$ denote the cases where the scalar ($a_0$) 
or pseudoscalar ($\pi$) meson is produced from the vacuum, respectively.
Accordingly, the decay amplitudes can be written as 
${\cal M}_{\rm SD}=(G_F/\sqrt{2})\,\lambda_d\, (T,C)$, 
where $G_F$ is the Fermi constant and $\lambda_q\equiv V_{cq}^*V_{uq}$ 
denotes the relevant Cabibbo--Kobayashi--Maskawa (CKM) matrix element. 
Using $a_2/a_1\sim -0.5$~\cite{Bajc:1997ey}, 
$f_{a_0}/f_\pi\sim 0.01$~\cite{Cheng:2022vbw,pdg}, and 
$F^{D^0\to \pi^-}/F^{D^{0}\to a_0^{-}}\sim 2.0$~\cite{Yang:2025gfz,Hsiao:2023qtk}, 
we obtain $r^{+/-}_{\rm SD}=0.065$ and $r^{+/0}_{\rm SD}=0.16$, 
as quoted in the introduction.

Even after including additional SD contributions, the discrepancy with the data remains unresolved. 
First, $T_{\rm S}(D^+\to\pi^0 a_0^{+})$ and $C_{\rm S}(D^+\to\pi^+ a_0^0)$ also contribute.
However, owing to the tiny scalar decay constant $f_{a_0}\sim 0.01\,f_\pi$,
these contributions are negligible compared to their counterparts
$C_{\rm P}(D^+\to\pi^0 a_0^+)$ and $T_{\rm P}(D^+\to\pi^+ a_0^{0})$, respectively.
Second, the $W$-annihilation ($W_{\rm an}$) and $W$-exchange ($W_{\rm ex}$) processes 
induce the transitions $D^{+}\to u\bar d\to\pi^{+(0)} a_0^{0(+)}$ and 
$D^{0}\to d\bar d\to \pi^\mp a_0^\pm$, respectively. 
In the $D$-meson rest frame, the quark pair $u(d)$ and $\bar d$ 
produced by the weak interaction move back-to-back with no relative orbital angular momentum. 
As a result, the $G$-parity of the $u\bar d$ ($d\bar d$) system 
can be identified with that of $\pi^+$ ($\pi^0$). 
Since $G(\pi)=-1$, whereas the final states satisfy $G(\pi a_0)=+1$, 
the $W_{\rm an}$ and $W_{\rm ex}$ contributions violate $G$-parity and
are therefore highly suppressed. As a result, they cannot enhance
$r^{+/-}_{\rm SD}$ and $r^{+/0}_{\rm SD}$. These shortcomings 
motivate the inclusion of additional topological configurations in the decay amplitudes, 
particularly those that favor the tetraquark interpretation of 
the $a_0$ meson~\cite{Achasov:2024nrh,Cheng:2024zul}.

%
\begin{figure}[t]
\includegraphics[width=2.8in]{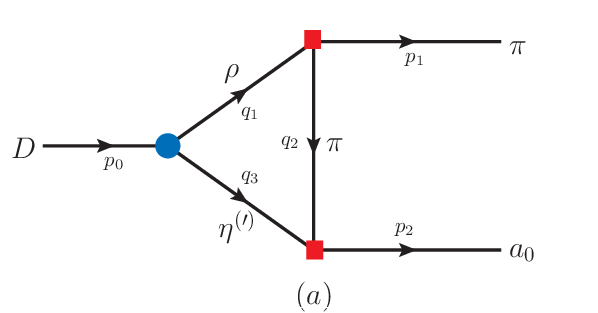}
\includegraphics[width=2.8in]{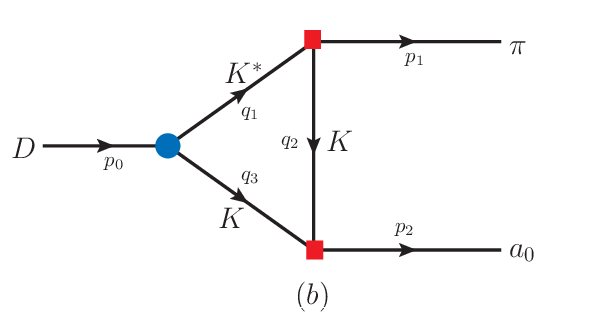}
\caption{Rescattering processes: (a) $D \to\rho\eta^{(\prime)}\to\pi a_0$ and
(b) $D\to K^* K\to\pi a_0$, where the exchange particles are $\pi$ and $K$, 
respectively.}\label{figDp}
\end{figure}
%

Indeed, the discrepancies in the measured ratios reflect the fact that 
the charm quark, with $m_c\simeq 1.3$~GeV, 
is not sufficiently heavy compared to the beauty quark, 
rendering heavy quark effective theory (HQET) less reliable 
for charmed hadron decays~\cite{Neubert:1993mb}. 
Consequently, nonperturbative contributions, particularly LD effects 
arising from final-state interactions or rescattering processes, 
play an essential role and must be properly incorporated.
In particular, the inclusion of such effects has been employed 
to resolve the discrepancies between the experimental measurements 
and short-distance expectations for the branching-fraction ratios and 
$\Delta {\cal A}_{CP}$ in $D^0\to \pi^+\pi^-$ and $D^0\to K^+ K^-$ decays~\cite{Bediaga:2022sxw,Cheng:2024hdo,Schacht:2021jaz,
Grossman:2019xcj,Franco:2012ck,Buccella:2019kpn}.

By reconciling theoretical predictions with experimental observations in 
$D_s^+\to \pi^+\omega$, 
$D_s^+\to\pi^{0}a_0^{+}(\pi^{+}a_0^{0})$, and
$D_s^+\to\rho^0 a_0^{+}$, it has been demonstrated that LD calculations  
can be broadly applied to non-leptonic $D\to PV$, $PS$, and $VS$ decays.
In particular, for the $D\to SS$ decay modes,
the channels $D_s^+\to\sigma_0 a_0^+$ and $D^+\to\sigma_0 a_0^+$,
with $\sigma_0\equiv f_0(500)$, receive negligible SD contributions, 
leading to naive expectations of ${\cal B}\sim 0$. However, LD rescattering effects 
enhance these rates to ${\cal B}\sim 1\times 10^{-2}$ and 
${\cal B}\sim 1\times 10^{-3}$~\cite{Wang:2025mdn}, respectively, 
in agreement with the subsequent experimental 
measurements~\cite{BESIII:2026mtz,BESIII:2026mbo}. 

Therefore, theoretical treatments of LD rescattering effects remain essential 
even for processes involving light scalar mesons, where neither the conventional 
$q\bar q$ nor the exotic tetraquark interpretations should be imposed \emph{a priori}.
Similarly, LD effects arising from rescattering processes can be incorporated in our analysis.
In Fig.~\ref{figDp}(a), the process $D\to \rho\eta^{(\prime)}\to\pi a_0$ 
originates from the SD weak decay $D\to \rho\eta^{(\prime)}$. 
The produced $\rho$ and $\eta^{(\prime)}$ mesons subsequently
undergo final-state interactions via pion exchange, 
rescattering into the $\pi$ and $a_0$ states.
In Fig.~\ref{figDp}(b), we also consider the rescattering process $D\to K^* K\to \pi a_0$, 
where the $K^* K\to \pi a_0$ transition proceeds through kaon exchange.

For clarity, we summarize the rescattering channels
$D\to V_1 P_2\to\pi a_0$ with $V_1 P_2=(\rho\eta^{(\prime)},K^* K)$ 
and the exchanged particle $P_3=(\pi,K)$, as follows
\begin{eqnarray}\label{Res1}
&&
D^0\to (\rho^0\eta^{(\prime)},K^{*-}K^+,K^{*0}\bar K^0)\to \pi^- a_0^+\,,\;
\text{with exchange of}~(\pi^+,\bar K^0,K^+)\,,\nonumber\\
&&
D^0\to (\rho^0\eta^{(\prime)},K^{*+}K^-,\bar K^{*0} K^0)\to \pi^+ a_0^-\,,\;
\text{with exchange of}~(\pi^-,K^0,K^-)\,,\nonumber\\
&&
D^+\to (\rho^+\eta^{(\prime)},K^{*+}\bar K^0,\bar K^{*0} K^+)\to \pi^0 a_0^+\,,\;
\text{with exchange of}~(\pi^+,K^+,\bar K^0)\,,\nonumber\\
&&
D^+\to (\rho^+\eta^{(\prime)},K^{*+}\bar K^0,\bar K^{*0} K^+)\to \pi^+ a_0^0\,,\;
\text{with exchange of}~(\pi^0,K^0,K^-)\,. 
\end{eqnarray}
The transition $D\to V_1P_2\to \pi a_0$ proceeds via a triangle loop, where
the numerator of the loop integral is constructed from the amplitudes
\begin{eqnarray}\label{LDamp}
&&{\cal M}(D\to V_1P_2)=\frac{G_F}{\sqrt 2}\lambda_q  C_{DV_1 P_2} \epsilon\cdot (p_0+q_2)\,,\nonumber\\
&&{\cal M}(V_1\to \pi P_3)=C_{V_1 P_3\pi} \epsilon\cdot (q_1-q_3)\,,\nonumber\\
&&{\cal M}(P_2\to a_0 P_3)=C_{a_0 P_2 P_3}\,.
\end{eqnarray}
Here, $C_{DV_1 P_2}$, $C_{V_1 P_3\pi}$, and $C_{a_0 P_2 P_3}$ 
denote the relevant coupling constants, $\epsilon$ is the polarization vector of the vector meson $V_1$, 
and $p_0$ and $(q_1,q_2,q_3)$ represent the momenta of the $D$ meson and $(V_1,P_2,P_3)$, 
respectively. The rescattering amplitude is then 
given by~\cite{Fajfer:2003ag,Hsiao:2019ait,Ling:2021qzl,Yu:2021euw}
\begin{eqnarray}\label{Mres}
&&{\cal M}_{\rm LD}^{(V_1 P_2)}\equiv
{\cal M}(D\to V_1 P_2\to\pi a_0)
=\frac{G_F}{\sqrt 2}\lambda_q \nonumber\\
&\times&C_{DV_1 P_2} C_{V_1 P_3\pi}C_{a_0 P_2 P_3}
\int \frac{d^4{q}_{1}}{(2\pi)^{4}}
\frac{(p_0+q_2)_{\mu}(-g^{\mu\nu}+\frac{q_1^{\mu}q_1^{\nu}}{q_1^2})(q_1-q_3)_{\nu}
F_{P_3}(q_{3}^2)}
{(q_1^2-m_1^2)(q_2^2-m_2^2)(q_3^2-m_3^2)}\,,
\end{eqnarray}
where $m_1$, $m_2$, and $m_3$ are the masses of the intermediate particles 
$V_1$, $P_2$, and the exchanged particle $P_3$, respectively. The form factor 
$F_{P_3}(q_{3}^2)\equiv (\Lambda_{P_3}^{2}-m_3^{2})/(\Lambda_{P_3}^{2}-q_{3}^{2})$, 
with the cutoff parameter $\Lambda_{P_3}$, 
is introduced to regularize the loop integral~\cite{Du:2021zdg}.
The total amplitude of $D\to \pi a_0$, by combining 
the SD and LD contributions in Eqs.~(\ref{TandC}) and (\ref{LDamp}), 
is given as
\begin{eqnarray}\label{ampT}
{\cal M}_T(D\to \pi a_0)={\cal M}_{\rm SD}
+{\cal M}_{\rm LD}^{(\rho\eta+\rho\eta')}e^{i\delta_1} 
+{\cal M}_{\rm LD}^{(K_1^* K_1+K_2^* K_2)}e^{i\delta_2}\,,
\end{eqnarray}
where $(\delta_1,\delta_2)$ denote the relative phases among the amplitudes, and 
can be determined in the numerical analysis.

\section{Numerical analysis}
%
\begin{table}[b]
\caption{Branching fractions for the initial $D\to V_1 P_2$ weak decays, 
together with the extracted coupling constants $C_{D V_1 P_2}$ 
(in units of $10^{-2}~{\rm GeV}^2$).}\label{tab1}
{
\scriptsize
\begin{tabular}
{|l|c|c|}
\hline
$D\to V_1 P_2$& branching fraction~\cite{pdg,CLEO:2008icw}
&$C_{DV_1 P_2}$\\
\hline\hline
$D^0\to\rho^0\eta,\rho^0\eta'$
&$(1.8\pm 0.3,4.5\pm 1.7)\times 10^{-4}$
&$(3.4\pm 0.3,14.4\pm 3.0)$
\\
$D^0\to K^{*-}K^+,K^{*0}\bar K^0$
&$(16.5\pm 1.1, 3.4\pm0.6)\times 10^{-4}$
&$(13.4\pm 0.4, 6.0\pm 0.6)$
\\
$D^0\to K^{*+}K^-,\bar K^{*0} K^0$
&$(45.6\pm 2.1, 2.5\pm 0.5)\times 10^{-4}$
&$(22.2\pm 0.5, 5.2\pm 0.5)$
\\
$D^+\to\rho^+\eta,\rho^+\eta'$
&$(1.9\pm 0.8,15.7\pm 5.0)\times 10^{-4}$
&$(2.3\pm 0.6,16.7\pm 3.0)$
\\
$D^+\to K^{*+}\bar K^0,\bar K^{*0} K^+$
&$(17.3\pm 1.8,3.7\pm0.2)\times 10^{-3}$
&$(27.4\pm 1.5,12.7\pm 0.3)$
\\
\hline
\end{tabular}
}
\end{table}
%
In the numerical analysis,  the CKM matrix elements in the Wolfenstein parameterization
are taken as
$V_{cd}=-\lambda+A^2 \lambda^5/2[1-2(\rho+i\eta)]$,
$V_{ud}=1-\lambda^2/2-\lambda^4/8$,
$V_{cs}=1-\lambda^2/2-\lambda^4/8(1+4A^2)$, and
$V_{us}=\lambda$~\cite{pdg},
with $(\lambda,A)=(0.225,0.826)$ and $(\rho,\eta)=(0.163\pm 0.010,0.357\pm 0.010)$.
For the SD amplitudes of $D\to \pi a_0$ in Eq.~(\ref{TandC}),
we adopt $(a_1,a_2)=(1.26\pm 0.04,-0.51\pm 0.05)$~\cite{Bajc:1997ey} and 
$f_{\pi^+}=\sqrt 2 f_{\pi^0}=130$~MeV~\cite{pdg}
as theoretical inputs. For the $D\to a_0$ form factors, we take
$F^{D^{0(+)}\to a_0^{-(+)}}=\sqrt 2 F^{D^+\to a_0^0}=0.44\pm 0.03$,
as determined from semileptonic $D$ decays~\cite{Hsiao:2023qtk}.
In this framework, treating the $a_0$ meson as an exotic tetraquark ($q^2\bar q^2$) state
leads to ${\cal B}(D^0\to a_0^- e^+\nu_e,a_0^-\to\pi^-\eta)=(0.8\pm 0.1\pm 0.1)\times 10^{-4}$,
in agreement with the experimental value 
$(0.86\pm 0.17\pm 0.05)\times 10^{-4}$~\cite{BESIII:2024zvp}.
With these inputs, the SD branching fractions 
${\cal B}_{\rm SD}(D\to \pi a_0)$ presented in Table~\ref{tab2} are obtained.

%
\begin{table}[b]
\caption{Branching fractions for $D\to \pi a_0$, together with the corresponding $CP$ asymmetries. 
The quoted uncertainties in our results, listed in order, arise from the cutoff parameters, 
the coupling constants, and the relative phases.}\label{tab2}
{
\scriptsize
\begin{tabular}
{|l|c|c|}
\hline
branching fraction and ${\cal A}_{CP}$& our work&data~\cite{BESIII:2024tpv}\\
\hline\hline
$10^4{\cal B}_{\rm SD}(D^0\to\pi^- a_0^+)$
&$\sim 0$
&
\\
$10^4{\cal B}(D^0\to(\rho^0\eta,\rho^0\eta')\to\pi^- a_0^+)$
&$(0.4\pm 0.0\pm 0.1, 2.2\pm 0.4\pm 1.0)$
&
\\
$10^4{\cal B}(D^0\to (K^{*-}K^+,K^{*0}\bar K^0)\to \pi^- a_0^+)$
&$(2.1\pm 0.5\pm 0.4, 0.4\pm 0.1\pm 0.1)$
&
\\
$10^4{\cal B}_T(D^0\to \pi^- a_0^+)$
&$(7.3\pm 0.9^{+1.1}_{-0.9}\pm 0.7)$
&$(8.3\pm 1.3)$
\\
$10^3 {\cal A}_{CP}(D^0\to \pi^- a_0^+)$
&$(-0.7\pm 0.1\pm 0.1\pm 0.1)$
& 
\\
\hline
$10^4{\cal B}_{\rm SD}(D^0\to\pi^+ a_0^-)$
&$(5.2\pm 1.2)$
&
\\
$10^4{\cal B}(D^0\to(\rho^0\eta,\rho^0\eta')\to\pi^+ a_0^-)$
&$(0.4\pm 0.0\pm 0.1, 2.2\pm 0.4\pm 1.0)$
&
\\
$10^4{\cal B}(D^0\to (K^{*+}K^-,\bar K^{*0} K^0)\to \pi^+ a_0^-)$
&$(5.9^{+1.5}_{-1.1}\pm 0.8, 0.3\pm 0.1\pm 0.1)$
&
\\
$10^4{\cal B}_T(D^0\to \pi^+ a_0^-)$
&$(1.2\pm 0.7\pm 0.7\pm 0.3)$
&$(1.1\pm 0.3)$
\\
$10^3 {\cal A}_{CP}(D^0\to \pi^+ a_0^-)$
&$(-2.1\pm 0.9\pm 1.1\pm 0.4)$
& 
\\
\hline\hline
$10^4{\cal B}_{\rm SD}(D^+\to\pi^0 a_0^+)$
&$(1.1\pm 0.4)$
&
\\
$10^4{\cal B}(D^+\to(\rho^+\eta,\rho^+\eta')\to \pi^0 a_0^+)$
&$(0.4\pm 0.0\pm 0.2,7.4\pm 1.4^{+3.0}_{-2.7})$
&
\\
$10^4{\cal B}(D^+\to (K^{*+}\bar K^0,\bar K^{*0} K^+)\to\pi^0 a_0^+)$
&$(11.1^{+2.8+2.3}_{-2.2-2.0},2.3^{+0.6+0.3}_{-0.4-0.1})$
&
\\
$10^4{\cal B}_T(D^+\to\pi^0 a_0^+)$
&$(14.3^{+2.5+1.9}_{-2.3-1.6}\pm 2.1)$
&$(14.4\pm 2.0)$
\\
$10^3 {\cal A}_{CP}(D^+\to\pi^0 a_0^+)$
&$(-1.4\pm 0.1\pm 0.1\pm 0.1)$
& 
\\
\hline
$10^4{\cal B}_{\rm SD}(D^+\to\pi^+ a_0^0)$
&$(6.4^{+1.3}_{-1.1})$
&
\\
$10^4{\cal B}(D^+\to(\rho^+\eta,\rho^+\eta')\to \pi^+ a_0^0)$
&$(0.4\pm 0.0\pm 0.2,7.4\pm 1.4^{+3.0}_{-2.7})$
&
\\
$10^4{\cal B}(D^+\to (K^{*+}\bar K^0,\bar K^{*0} K^+)\to\pi^+ a_0^0)$
&$(11.1^{+2.8+2.3}_{-2.2-2.0},2.3^{+0.6+0.3}_{-0.4-0.1})$
&
\\
$10^4{\cal B}_T(D^+\to\pi^+ a_0^0)$
&$(5.5^{+2.4+2.6}_{-2.0-2.3}\pm 0.6)$
&$(5.6\pm 1.6)$
\\
$10^3 {\cal A}_{CP}(D^+\to\pi^+ a_0^0)$
&$(0.2\pm 0.0\pm 0.1\pm 0.1)$
& 
\\
\hline
\end{tabular}
}
\end{table}
%

The evaluation of the LD contributions requires the coupling constants $C_{D V_1 P_2}$, 
which are extracted from the measured branching fractions of the $D\to V_1 P_2$ decays. 
The corresponding values are listed in Table~\ref{tab1}.
In particular, the decay mode $D^{0(+)}\to \rho^{0(+)}\eta'$ 
has not been directly measured, but can be inferred from the resonant process 
$D^{0(+)}\to \pi^+\pi^{-(0)}\eta'$~\cite{CLEO:2008icw}. 
Since this channel is dominated by the sequential decay
$D^{0(+)}\to \rho^{0(+)}\eta'$ followed by $\rho^{0(+)}\to \pi^+\pi^{-(0)}$, 
we employ the approximation ${\cal B}(D^{0(+)}\to\pi^+\pi^{-(0)}\eta')\simeq 
{\cal B}(\rho^{0(+)}\to \pi^{0(+)}\pi^-)\times {\cal B}(D^{0(+)}\to\rho^{0(+)}\eta')$.
We thus extract ${\cal B}(D^{0(+)}\to \rho^{0(+)}\eta')$, which are found to be 
consistent with the theoretical results in Ref.~\cite{Zheng:2025ryf}.

For the strong interaction vertices, the coupling constants are determined using experimental inputs. 
In particular, we take $C_{\rho \pi\pi}=6.0$, and
$C_{K^{*+} K^0\pi^+}/\sqrt 2 =C_{K^{*+} K^+\pi^0}=3.23$
($C_{K^{*0} K^+\pi^-}/\sqrt 2=C_{K^{*0} K^0\pi^0}=3.14$),
which are obtained from the branching fractions ${\cal B}(\rho^{0(+)}\to \pi^+\pi^{-(0)})\simeq 100\%$ 
and ${\cal B}(K^{+(0)}\to K^{0(-)}\pi^+)=2{\cal B}(K^{+(0)}\to K^{+(0)}\pi^0)=2/3$~\cite{pdg}.
For the scalar couplings $C_{a_0 P_2 P_3}$, we adopt the values extracted 
in Refs.~\cite{Bugg:2008ig,BESIII:2016tqo}, namely
$(C_{a_0\pi\eta},C_{a_0\pi\eta'})=(2.87\pm 0.09,2.52\pm 0.08)$~GeV for
$a_0^{+(0)}\to \pi^{+(0)}\eta$ and $a_0^{+(0)}\to \pi^{+(0)}\eta'$, respectively, and
$C_{a_0 KK}=(2.94\pm 0.13),\text{GeV}$, defined through
${\cal M}(a_0^+\to K^+\bar K^0)=(-)\sqrt 2{\cal M}(a_0^0\to K^+ K^-(K^0\bar K^0))=C_{a_0 KK}$.

The cutoff parameters are taken as $(\Lambda_\pi,\Lambda_K)=(1.2\pm 0.3, 1.5\pm 0.3),\text{GeV}$, 
subject to the constraint $\Lambda_K-\Lambda_\pi=m_K-m_\pi$~\cite{Cheng:2004ru}.
The variation $\delta \Lambda_{\pi(K)}=0.3$~GeV 
is introduced to estimate the sensitivity of the loop calculations.
Notably, $\Lambda_\pi\simeq 1$~GeV has been shown to be a universal cutoff scale
in describing hadronic rescattering contributions to two-body $D$ decays~\cite{Tornqvist:1993ng,
Li:1996yn,Wu:2019vbk,Yu:2021euw,Hsiao:2019ait,Wang:2025mdn}.
Incorporating these coupling constants and cutoff parameters 
into the integration of the rescattering amplitude in Eq.~(\ref{Mres}), 
we obtain the corresponding rescattering branching fractions,
as presented in Table~\ref{tab2}.

The resonant three-body decays $D\to \pi a_0,a_0\to \pi\eta$ 
have been experimentally observed, whereas the corresponding two-body decays 
$D\to \pi a_0$ are investigated in this work. 
To relate the measured decay chain $D\to \pi a_0,a_0\to \pi\eta$ 
to its underlying two-body subprocess, 
we employ the relation ${\cal B}(D\to \pi a_0,a_0\to\pi\eta)=F_{\rm RE}{\cal B}(D \to \pi a_0)$,
where the factor $F_{\rm RE}$ accounts for the resonant contribution of 
$a_0\to \eta\pi$ and is estimated to be $0.66$ in Ref.~\cite{BESIII:2019jjr}.
Using the measured branching fractions of $D\to \pi a_0,a_0\to \pi\eta$~\cite{BESIII:2024tpv}, 
we extract ${\cal B}_{\rm ex}(D\to \pi\eta)$, 
as given in Table~\ref{tab2}. 

On the theoretical side, the total branching fractions ${\cal B}_T(D\to \pi\eta)$ 
are calculated from the total amplitude in Eq.~(\ref{ampT}). 
By imposing the condition ${\cal B}_T(D\to \pi\eta)={\cal B}_{\rm ex}(D\to \pi\eta)$, 
we obtain four constraint equations, whose solutions lead to
\begin{eqnarray}\label{phase2}
(\delta_1^0,\delta_2^0)&=&(-163.3\pm 4.3,-50.5\pm 4.2)^\circ\,,\nonumber\\
(\delta_1^+,\delta_2^+)&=&(-98.2\pm 4.3,45.3\pm 3.6)^\circ\,,
\end{eqnarray}
where the phases $\delta_{1,2}^{0}$ and $\delta_{1,2}^{+}$ 
correspond to the $D^0$ and $D^+$ decay modes, respectively, 
and provide essential strong-phase inputs for the direct $CP$ asymmetry.
Using the definition
\begin{eqnarray}\label{Acp}
{\cal A}_{CP}(D\to \pi a_0)=\frac{{\cal B}(D\to \pi a_0)-{\cal B}(\bar D\to \bar\pi \bar a_0)}
{{\cal B}(D\to \pi a_0)+{\cal B}(\bar D\to \bar\pi \bar a_0)}\,,
\end{eqnarray}
we evaluate ${\cal A}_{CP}(D\to \pi a_0)$,
where $\bar D\to \bar\pi \bar a_0$ denotes the $CP$-conjugate process of $D\to \pi a_0$.
The resulting ${\cal B}_T(D\to \pi a_0)$ and ${\cal A}_{CP}(D\to \pi a_0)$ 
are summarized in Table~\ref{tab2}.

\section{Discussions and Conclusion}
The SD predictions for ${\cal B}_{\rm SD}(D^0\to\pi^- a_0^+,\pi^+ a_0^-)$ and
${\cal B}_{\rm SD}(D^+\to\pi^0 a_0^+)$ are clearly inconsistent with the experimental data.
In particular, ${\cal B}_{\rm SD}(D^0\to\pi^- a_0^+)\sim 0$.
Once the LD rescattering effects are included, 
a dominant contribution ${\cal B}_{\rm LD}(D^0\to(\rho^0\eta'+K^{*-}K^+)\to\pi^- a_0^+)
\simeq 4\times 10^{-4}$ is obtained, yielding a non-vanishing contribution. 
Moreover, constructive interference arising from the relative phases 
$\delta_1^0$ and $\delta_2^0$ in Eq.~(\ref{LDamp})
further enhance the total branching fraction, leading to
${\cal B}_T(D^0\to\pi^- a_0^+)\simeq {\cal B}_{\rm ex}(D^0\to\pi^- a_0^+)$.
Remarkably, this result corresponds to a significance of $5.0\sigma$.

From Table~\ref{tab2}, ${\cal B}_{\rm SD}(D^0\to\pi^+ a_0^-)$,
${\cal B}(D^0\to \rho^0\eta'\to\pi^+ a_0^-)$, and
${\cal B}(D^0\to K^{*+}K^-\to \pi^+ a_0^-)$ are approximately 
5, 2, and 5 times larger than the measured value. 
When both SD and LD contributions are taken into account, 
the same set of the relative phases ($\delta_1^0$ and $\delta_2^0$) 
lead to destructive interference. As a result,
${\cal B}_T(D^0\to \pi^+ a_0^-)$ is reduced to a value 
consistent with the experimental data.
Nonetheless, these sizable contributions propagate relatively large uncertainties, 
especially given the reduced central value. 

The inclusion of SD and LD contributions, 
together with their interference, brings ${\cal B}_T(D^+\to\pi^0 a_0^+)$ and 
${\cal B}_T(D^+\to\pi^+ a_0^0)$ into agreement with the data. In particular, since
${\cal B}_{\rm SD}(D^+\to\pi^+ a_0^0)$, ${\cal B}(D^+\to \rho^+\eta'\to \pi^+ a_0^0)$, and
${\cal B}(D^+\to (K^{*+}\bar K^0,\bar K^{*0} K^+)\to\pi^+ a_0^0)$ are comparable to
or exceed the experimental value, destructive interference is required, thereby reducing
${\cal B}_T(D^+\to\pi^+ a_0^0)$. As a consequence, the propagated uncertainties 
become comparable to the central value. In contrast, in the absence of strong cancellations,
the uncertainties of ${\cal B}_T(D^+\to\pi^0 a_0^+)$ remain well controlled,
corresponding to a significance of $4.1\sigma$.

With the amplitude written as 
${\cal M}=\lambda_d |{\cal M}_d|^{i\delta_d}+\lambda_s |{\cal M}_s|e^{i\delta_s}$,
the direct $CP$ asymmetry in Eq.~(\ref{Acp}) 
can be recast as
\begin{eqnarray}\label{Acp2}
{\cal A}_{CP}\simeq 
\frac{F_{\rm W}R_{\rm M}\sin\Delta}{1+R_{\rm M}^2-2R_{\rm M}\cos\Delta}\,,
\end{eqnarray}
where, for SCS $D$ decays, we have used the approximation $\lambda_s\simeq -\lambda_d$. 
We define $F_{\rm W}=-2\,{\rm Im}(\lambda_d\lambda_s^*)/|\lambda_d|^2$,
$R_{\rm M}=|{\cal M}_s|/|{\cal M}_d|$, and $\Delta=\delta_d-\delta_s$.
The value $F_{\rm W}\simeq -1.3\times 10^{-3}$ reflects the small weak phase 
originating from the CKM parameters, 
thereby serving as an intrinsic suppression factor for ${\cal A}_{CP}$.

In the SCS $D\to\pi a_0$ decays, the SD amplitude contributes only to ${\cal M}_d$,
and thus cannot by itself generate a nonzero $CP$ asymmetry.
A non-vanishing ${\cal A}_{CP}$ arises once additional contributions to ${\cal M}_s$
are included, such as those from rescattering processes
$D\to (K_1^* K_1+K_2^* K_2)\to\pi  a_0$. Moreover, the strong phases
$\delta_1$ and $\delta_2$, together with the absorptive phases generated by
integrating over the on-shell regions of the propagators, 
provide significant contributions to the total phases $\delta_d$ and $\delta_s$.

Numerically, we obtain $R_{\rm M}=(1.1,1.4,1.1,2.7)$ and
$\Delta=(83.5,17.0,48.8,-131.9)^\circ$, which correspond to the predictions
${\cal A}_{CP}(D^0\to\pi^- a_0^+)=(-0.7\pm 0.1\pm 0.1\pm 0.1)\times 10^{-3}$, 
${\cal A}_{CP}(D^0\to\pi^+ a_0^-)=(-2.1\pm 0.9\pm 1.1\pm 0.4)\times 10^{-3}$, 
${\cal A}_{CP}(D^+\to\pi^0 a_0^+)=(-1.4\pm 0.1\pm 0.1\pm 0.1)\times 10^{-3}$, and 
${\cal A}_{CP}(D^+\to\pi^+ a_0^0)=(0.2\pm 0.0\pm 0.1\pm 0.1)\times 10^{-3}$, respectively.
Clearly, $R_{\rm M}\simeq 1$ indicates comparable magnitudes of 
$|{\cal M}_s|$ and $|{\cal M}_d|$, leading to ${\cal A}_{CP}$ at the $10^{-3}$ level.
In contrast, for $R_{\rm M}\simeq 2.7$, the imbalance between $|{\cal M}_s|$ and $|{\cal M}_d|$ 
suppresses ${\cal A}_{CP}(D^+\to\pi^+ a_0^0)$ to the $10^{-4}$ level.
%
\begin{figure}[t]
\includegraphics[width=2in]{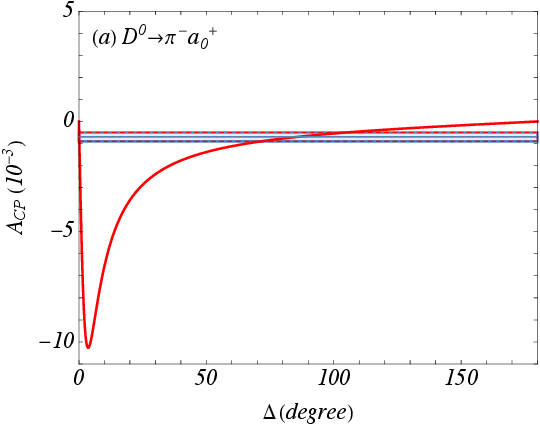}
\includegraphics[width=1.95in]{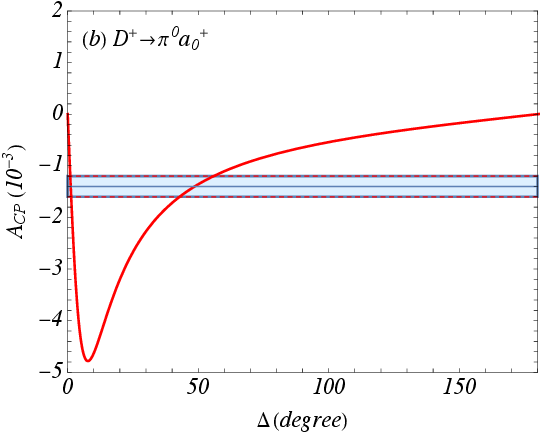}
\caption{$CP$ asymmetries for (a)~$D^0\to \pi^- a_0^+$ and (b)~$D^+\to \pi^0 a_0^+$ 
as functions of the relative phase $\Delta$ defined in Eq.~(\ref{Acp2}). The shaded regions indicate 
the central values and their uncertainties obtained from the numerical analysis.}\label{figAcp}
\end{figure}
%

Owing to the absence of severe destructive interference for the branching fractions,
the resulting ${\cal A}_{CP}(D^0\to\pi^- a_0^+)$ and ${\cal A}_{CP}(D^+\to\pi^0 a_0^+)$ 
are well controlled, with small uncertainties. The relative phase $\Delta$ also plays a crucial role
in determining their magnitudes. As illustrated in Fig.~\ref{figAcp}(a), 
${\cal A}_{CP}(D^0\to\pi^- a_0^+)$ depends sensitively on $\Delta$.
In particular, $\Delta=0^\circ$ or $180^\circ$ leads to ${\cal A}_{CP}=0$, 
while varying $\Delta$ from $180^\circ$ to $360^\circ$ results in a sign flip.
At $\Delta\simeq 4^\circ$, ${\cal A}_{CP}$ reaches 
an extreme value of about $-10\times 10^{-3}$. Similarly, Fig.~\ref{figAcp}(b) shows that 
${\cal A}_{CP}(D^+\to\pi^0 a_0^+)$ exhibits a comparable sensitivity.
Our numerical results are represented by the shaded regions, 
which indicate the central values together with their uncertainties 
for ${\cal A}_{CP}(D^0\to\pi^- a_0^+)$ and ${\cal A}_{CP}(D^+\to\pi^0 a_0^+)$.
Notably, the allowed ranges of $\Delta$ do not overlap with 
those corresponding to extreme cancellations or enhancements. 

As a final remark, the LD rescattering effects required in SCS $D\to \pi a_0$ decays 
can be naturally extended to more general SCS $D\to PS$ and $D\to VS$ modes.
Notably, the rescattering-induced contributions can lead to $|{\cal M}_s|\sim |{\cal M}_d|$, 
thereby enhancing the resulting $CP$ asymmetries to the ${\cal O}(10^{-3})$ level.
This provides a novel testing ground for probing $CP$ violation in the charm sector 
without invoking penguin contributions or physics beyond the SM.

In summary, we have performed a systematic investigation of the SCS $D\to \pi a_0$ decays.
We have shown that the SD amplitudes alone are far from sufficient to account for 
the observed branching-fraction ratios, thereby indicating the necessity of LD rescattering effects.
In particular, the processes $D\to \rho\eta^{(\prime)}\to \pi a_0$ and 
$D\to K^*K\to \pi a_0$ have provided sizable contributions, with the latter 
playing a dominant role in generating ${\cal M}_s$ in the amplitude 
${\cal M}=\lambda_d{\cal M}_d+\lambda_s{\cal M}_s$, accompanied by nontrivial 
strong phases that are essential for direct $CP$ asymmetries.
We have thus predicted ${\cal A}_{CP}(D^0\to\pi^- a_0^+,\pi^+ a_0^-)
=(-0.7\pm 0.1\pm 0.1\pm 0.1,\,-2.1\pm 0.9\pm 1.1\pm 0.4)\times 10^{-3}$ and 
${\cal A}_{CP}(D^+\to\pi^0 a_0^+)=(-1.4\pm 0.1\pm 0.1\pm 0.1)\times 10^{-3}$, 
which naturally reach the ${\cal O}(10^{-3})$ level and are accessible 
to experiments at BESIII, Belle~II, and LHCb.
These results open a new avenue for probing $CP$ violation in SCS $D\to \pi a_0$ decays.

\section*{ACKNOWLEDGMENTS}
The authors would like to thank Dr.~Yu Lu for valuable discussions.
This work was supported in part by the National Natural Science Foundation of China 
(Grants~No.~12575101 and No.~12175128).


\end{document}